\documentclass[reprint,superscriptaddress,showpacs,preprintnumbers,nofootinbib,amsmath,amssymb,floatfix,aps]{revtex4-1}
\usepackage{graphicx}% Include figure files
\usepackage{subfigure}%
\usepackage{isotope}

\newcommand{\ve}[1]{\ensuremath{\mathbf{#1}}}
\newcommand{\n}[1]{\ensuremath{|\mathbf{#1}|}}
\newcommand{\xsec}{\ensuremath{d\sigma/d\omega d\Omega}}
\newcommand{\Erec}{\ensuremath{E_\nu^\textrm{rec}}}
\newcommand{\Emax}{\ensuremath{E_\nu^\textrm{max}}}
\newcommand{\etal}{{\it et al.}}
\newcommand{\Ek}{  \ensuremath{ E_{\mathbf k} }  }
\newcommand{\tk}{  \ensuremath{ t_\textrm{kin} }  }
\newcommand{\sFSI}{  \ensuremath{ \sigma^\textrm{FSI} }  }
\newcommand{\sIA}{  \ensuremath{ \sigma^\textrm{IA} }  }

\newcommand{\Epp}{\ensuremath{E_{\mathbf p'}}}

\graphicspath{{figs/}}

\begin{document}

\title{Improving the accuracy of neutrino energy reconstruction\\ in charged-current quasielastic scattering off nuclear targets}
\author{Artur M. Ankowski}
\altaffiliation{Present address: Center for Neutrino Physics, Virginia Tech, Blacksburg, Virginia 24061, USA}
\email{artank@vt.edu}
\affiliation{Department of Physics, Okayama University, Okayama 700-8530, Japan}
\author{Omar Benhar}
%\email{Omar.Benhar@roma1.infn.it}
\affiliation{INFN and Department of Physics,``Sapienza'' Universit\`a di Roma, I-00185 Roma, Italy}
\author{Makoto Sakuda}
%\email{Sakuda@fphy.hep.okayama-u.ac.jp}
\affiliation{Department of Physics, Okayama University, Okayama 700-8530, Japan}

\date{\today}%

\begin{abstract}
We report the results of a theoretical study of quasielastic electron and  neutrino interactions with carbon.
Our approach takes into account the effects of final-state interactions between the struck nucleon and the residual
nucleus, neglected in the impulse approximation, through a generalization of the spectral function formalism.
The calculated electron-scattering cross sections turn out to be in very good agreement with the available data over
a broad kinematical region. The impact of nuclear effects on the reconstruction of neutrino energy
in charged-current quasielastic processes is also studied, and the results of our approach are compared to the predictions of the relativistic Fermi gas model, routinely employed in most Monte Carlo simulations. Finally, we discuss the limitations of the existing procedure for energy reconstruction and propose a new, improved, one.
At energy $\sim$600 MeV, we observe a sizable difference between neutrino and antineutrino scattering, important for the measurements of charge-parity symmetry violation.
Our analysis suggests that a reliable determination of neutrino energy can only be obtained from models
validated by a systematic comparison to the available electron-scattering data.

\end{abstract}

%13. Specific reactions and phenomenology
%13.15.+g 	Neutrino interactions
%
%25. Nuclear reactions: specific reactions
%25.30.Pt 	Neutrino-induced reactions
%25.30.Fj 	Inelastic electron scattering to continuum

%\keywords{Suggested keywords}%Use showkeys class option if keyword display desired

\maketitle

\section{Introduction}

The results of neutrino experiments reported in the past few years have significantly improved our knowledge of the oscillation parameters~\cite{ref:Fogli,ref:Forero}. Further progress, including constraints on the neutrino-mass hierarchy and the phase violating charge-parity (CP) symmetry, is expected to come from ongoing and future measurements by, for example, the T2K~\cite{ref:T2K_oscillations} and NO$\nu$A~\cite{ref:NOvA} Collaborations. However, oscillation studies carried out with accelerator and atmospheric neutrinos heavily rely on the reconstruction of neutrino energy. Therefore, the correct interpretation of their outcome requires an accurate estimate of the neutrino cross sections for the relevant nuclear targets.

The description of charged-current (CC) quasielastic (QE) scattering plays a particularly important role, since this reaction mechanism is known to be dominant at neutrino energy $\sim$600 MeV, the kinematical setup of T2K, and it yields a sizable contribution to the total cross section in the few-GeV region~\cite{ref:delta}, the kinematics of NO$\nu$A.

Owing to the difficulties of hadron reconstruction, in CC QE events, the neutrino energy is typically reconstructed from the measured kinematics of the charged lepton only. The accuracy of this method is limited by the accuracy to which nuclear effects are described by the Monte Carlo simulations involved in data analysis.

The simulations of many recent experiments~\cite{ref:MiniB_kappa,ref:MINOS_QE,ref:NOMAD} were performed using the relativistic Fermi gas (RFG) model, with different {\it ad hoc} modifications introduced as a remedy for its shortcomings. The development of more realistic models of nuclear structure and dynamics, capable to provide a truly quantitative description of neutrino cross sections,  appears to be needed to meet the requirements of precise neutrino oscillation studies.

In this article, we develop an approach suitable to provide an estimate of the CC QE cross section across the broad range of neutrino energy, $0.1\lesssim E_\nu\lesssim10$~GeV, relevant to accelerator and atmospheric neutrino experiments.
We employ a generalization of the impulse approximation (IA) scheme~\cite{ref:Omar_oxygen,ref:Omar_FSI_NM}, based on the assumption that the interaction between the beam particle and the nucleus involves a single nucleon, the remaining $(A-1)$ nucleons acting as spectators. Within this picture, the information on the target initial state is contained in the {\em hole} spectral function (SF), while the propagation of the struck nucleon in the final state is described by the {\em particle} SF.
Our calculations make use of the hole SF obtained in Ref.~\cite{ref:Omar_LDA} combining experimental information and accurate many-body calculations.
Final-state interactions (FSI) between struck nucleon and the spectator system are taken into account within the correlated Glauber approximation discussed in Refs.~\cite{ref:Omar_FSI,ref:Omar_FSI_NM}.

The main original feature of our work is the inclusion, in the energy spectrum of the struck nucleon, of the real part of the optical potential obtained from the
Dirac phenomenological analysis of Cooper \etal{}~\cite{ref:Cooper_EDAI}. We consider the carbon nucleus, employed in the NO$\nu$A detectors~\cite{ref:NOvA} and in the T2K near detector~\cite{ref:T2K_CC}, and  restrict our discussion to the region of momentum transfer, $150\lesssim\n q\lesssim 500$~MeV, where the QE peak can be unambiguously identified.
Also, of paramount importance is the comparison to precise electron-scattering data~\cite{ref:Barreau,ref:Baran,ref:Whitney,ref:QES_archive}, which allows us to understand and quantify the uncertainties of our calculations, involving no adjustable parameters.

The strong motivation of our work becomes fully apparent when one considers that a firm handle on the uncertainties associated with the calculated cross sections translates into well-controlled uncertainties in neutrino energy reconstruction. To gauge the dependence of the unfolding of the neutrino cross section on the description of the scattering process, we compare the results obtained from our approach to the predictions of the RFG model. We also analyze the impact of FSI on energy reconstruction, and pay special attention to the difference between neutrino and antineutrino interactions arising from Coulomb effects. Finally, we discuss limitations of the standard energy reconstruction method,
and propose possible improvements to it.

As the problems of energy reconstruction and unfolding are essential for neutrino oscillation studies, their various aspects have been discussed in the literature. In their study  of neutrino-oxygen interactions, performed within an approach similar to ours, Benhar and Meloni~\cite{ref:Omar&Davide} observed that the unfolding procedure is affected by nuclear effects not taken into account within the RFG model. Leitner and Mosel~\cite{ref:Leitner} used the Giessen Boltzmann-Uehling-Uhlenbeck (GiBUU) transport model to analyze the effect of absorbed or undetected pions in carbon. Their results show that the reconstructed energy of such events is typically lower than the true one by $\sim$300 MeV. Martini {\etal}~\cite{ref:Martini_Erec} argued that multinucleon final states tend to redistribute the strength of the reconstructed flux from the peak to the tails, and that neglecting this effect in oscillation analysis affects the extracted oscillation parameters. Nieves {\etal}~\cite{ref:Nieves_Erec} showed that the unfolding of the total CC QE cross section performed without accounting for multinucleon effects on reconstructed energy may significantly distort its energy dependence. Finally, Lalakulich {\etal} \cite{ref:Lalakulich_Erec} extended the GiBUU model by adding the two-nucleon knockout contribution to the CC QE cross section, obtained from a fit to the MiniBooNE data based on a physically well-motivated  {\it ansatz}. They observed that, while the effect of two-nucleon processes on energy reconstruction is more relevant in the low-energy region, the role of pion-related backgrounds is more pronounced at higher energy.

Recently, Coloma and Huber~\cite{ref:Coloma_PRL} have reported a~sizable bias from nuclear effects in the determination of the oscillation parameters,
resulting from an analysis carried out using the extended GiBUU model. Coloma {\etal}~\cite{ref:Coloma_PRD} have also performed a comparison between the GiBUU and
GENIE Monte Carlo generators, used in data analysis by various experiments, and observed that while the CC QE results obtained without FSI effects are in good agreement, their inclusion leads to an apparent shift between the event distributions as a function of reconstructed energy.

To address the problem of FSI in CC QE interactions, we first deduce the accuracy our approach from comparisons to the precise $\isotope[12][6]{C}(e,e')$ cross sections and then analyze its predictions for neutrino and antineutrino interactions. We make use of the carbon optical potential of Ref.~\cite{ref:Cooper_EDAI}. The same potential is employed in the relativistic mean-field models of Refs.~\cite{ref:Udias,ref:Meucci,ref:Meucci_RGF}, extensively applied in analyses of electron and neutrino scattering~\cite{ref:Alberico_ROP,ref:Maieron,ref:Meucci_CCQE,ref:Martinez,ref:RMF,ref:RGF_CC,ref:Meucci_anu,ref:Meucci_MINERvA}.

In the mean-field models, FSI are accounted for by strong potentials and they have a significant effect, yielding the dominant contribution to the tail of the cross section at large energy transfer. Within the SF approach, on the other hand, the effect of FSI is much weaker and it leads to a slight enhancement of the cross section's tail originating from
short-range correlations between nucleons in the \emph{initial} state~\cite{ref:Omar_oxygen}.

This article is organized as follows.
The elements of our approach are outlined in Sec.~\ref{sec:Formalism}, while Sec.~\ref{sec:NumericalResults} is devoted to the discussion of numerical results. Finally, in Sec.~\ref{sec:Conclusions}
we state the conclusions.

\section{Formalism}\label{sec:Formalism}

In our approach, the cross section is obtained within the convolution scheme~\cite{ref:Omar_RMP}, which amounts to integrating the IA prediction with a folding function that describes the effects of FSI between the struck particle and the spectator system. The resulting expression is
\begin{equation}\label{eq:xsec_FSI}
\frac{d\sFSI}{d\omega d\Omega }= \int d\omega'f_{\ve{q}}(\omega-\omega')\frac{d\sIA}{d\omega' d\Omega},
\end{equation}
where $\omega$ and $\Omega$ are the energy transfer and the solid angle specifying the direction of the outgoing lepton, respectively.

The folding function can be decomposed in the form
\begin{equation}\label{eq:FF}
f_{\ve{q}}(\omega)=\delta(\omega)\sqrt{T_A}+\big(1-\sqrt{T_A}\big)F_{\ve{q}}(\omega),
\end{equation}
showing that the strength of FSI is driven by (i) the nuclear transparency $T_A$,  and (ii) the finite-width function $F_{\ve{q}}(\omega)$. Both these quantities are strongly affected by short-range nucleon-nucleon correlations, leading to a~suppression of the probability that
the struck nucleon interacts with the spectator particles within a distance $\lesssim1$~fm of the interaction vertex.

In the limit of  full  nuclear transparency to  the struck nucleon, i.e. $T_A\to 1$, the folding function reduces to $\delta(\omega)$, and the IA cross section is
recovered from Eq.~\eqref{eq:xsec_FSI}.

The effects of FSI on the cross section can be qualitatively understood in terms of the optical potential $U=U_V+iU_W$, originally proposed in the context of $(e,e')$ processes by Horikawa {\it et al.}~\cite{ref:Horikawa}. Within this picture, the real part of the potential, $U_V$, determines a modification of the energy spectrum of the final-state nucleon, while the imaginary part, $U_W$, moving strength from
single-particle excitations to more complex final states, is related to the folding function of Eq.\eqref{eq:FF} through
\[
F_{\ve{q}}(\omega)=\frac{1}{\pi}\frac{U_W}{U_W^2+\omega^2}.
\]
As a consequence, $U_V$ produces a shift of the cross section, while $U_W$  brings about a quenching of the QE
peak and the associated enhancement of its tails. Note, however, that in the optical potential model the latter effect is overestimated, owing to the infinite
tails of $F_{\ve{q}}(\omega)$~\cite{ref:Omar_FSI}.

To account for the modification of the struck nucleon's energy, we include $U_V$ in the argument of the folding function, replacing
\begin{equation}\label{eq:rOP}
f_{\ve{q}}(\omega-\omega')\to f_{\ve{q}}(\omega-\omega'-U_V).
\end{equation}

The above prescription is somewhat reminiscent of the procedure used in the Fermi gas model, in which an average nucleon-separation energy $\varepsilon$ is included in the argument of the energy-conserving $\delta$ function.

The proton optical potential of carbon has been determined by Cooper \etal{}~\cite{ref:Cooper_EDAI} using Dirac phenomenology. Within this approach, widely employed in analyses of electron-induced proton knockout and nucleon scattering~\cite{ref:pKnockout,ref:pInduced,ref:pScattering}, the optical potential is described by means of the (complex) scalar and vector potentials, $S$ and $V$, appearing in the Dirac equation. Their dependence on kinetic energy, $\tk$, and radial coordinate, $r$, is found by fitting the scattering solutions to the measured
elastic cross section, analyzing power, and spin rotation function, available for protons of kinetic energy in the range $29\leq \tk\leq1040$~MeV.

\begin{figure}[t,r]
\centering
    \includegraphics[width=0.85\columnwidth]{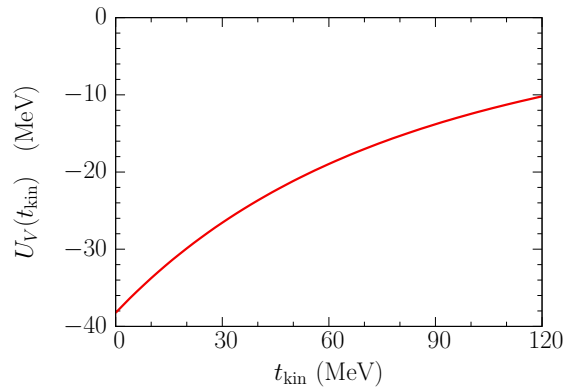}
\caption{\label{fig:potential}(color online). Real part of the carbon optical potential for proton, obtained from the Dirac phenomenological fit of Cooper \etal{}~\cite{ref:Cooper_EDAI}, as a function of proton's kinetic energy. }
\end{figure}

In the presence of the scalar and vector potentials, the total energy of proton $E'_\textrm{tot}=E'_\textrm{tot}(\tk,r)$ can be written in the form
\begin{equation}
E'_\textrm{tot}=\sqrt{(M+S)^2+\ve p'^2}+V,
\end{equation}
with $M$ and $\ve p'$ being the nucleon's mass and momentum, respectively. Because in our calculations the optical potential is an $r$-independent modification to the on-shell energy, $\Epp=\sqrt{M^2+\ve p'^2}$, it is simply related to $E'_\textrm{tot}$ through
\begin{equation}
\int  d^3r\rho(r) E'_\textrm{tot}=\Epp+U,
\end{equation}
where $\rho(r)$ denotes the nuclear density distribution. Hence, its real part is given by
\begin{equation}
U_V=\int  d^3r\rho(r)\Re(E'_\textrm{tot})-\Epp,
\end{equation}
where $\Re(x+iy)=x$. Using the density distribution of carbon---unfolded from the measured charge density~\cite{ref:densityDistributions} following to the procedure described in Ref.~\cite{ref:densityUnfolding}---and the $A$-independent fit of Ref.~\cite{ref:Cooper_EDAI}, we obtain the proton $U_V$ shown in Fig.~\ref{fig:potential}. It clearly appears that in the low-$\tk$ region, particularly relevant to QE scattering, interactions with the spectator system lead to a~sizable modification to the struck protons's spectrum. We assume that the neutron $U_V(\tk)$  only differs from the proton one
due to the (constant) Coulomb correction, which we estimate to be 3.5~MeV.

To evaluate the folding function~\eqref{eq:FF}, we use the nuclear transparency of carbon reported in Ref.~\cite{ref:Rohe}, and neglect the $\n q$ dependence
of $F_{\ve{q}}(\omega)$.
This choice is motivated by the results of Ref.~\cite{ref:Omar_RMP}, suggesting that, at large $\n q$,  $F_{\ve{q}}(\omega)$ depends weakly on momentum transfer.
In addition, its inclusion has a small effect---not exceeding 13\%---on the cross sections discussed in this paper.
The numerical results reported in this work are obtained with $F_{\ve{q}}(\omega)$ calculated at $\n q = 1$ GeV.

Note that in Eq.~\eqref{eq:xsec_FSI}, the nucleon kinematics is integrated out. Therefore, in our approach, $T_A=T_A(\tk)$ and $U_V=U_V(\tk)$ are evaluated at
\begin{equation}\label{eq:t}
\tk=\frac{\Ek^2(1 - \cos\theta)}{ M + \Ek(1 - \cos\theta)},
\end{equation}
where $\Ek$ and $\theta$ denote the energy of the beam particle and the angle of the outgoing lepton, respectively.
The above equation corresponds to scattering of a massless particle on a nucleon at rest.

The proton and neutron $(N=p\,,n)$ contributions to the IA cross section~\cite{ref:Omar_FSI_NM} are obtained from
\begin{eqnarray}
\frac{d\sIA_{\ell N}}{d\omega d\Omega }&=& \int d^3p\,
dEP^N_\textrm{hole}(\ve p,E)\frac{M}{E_{\ve p}}\frac{d\sigma^\textrm{elem}_{\ell N}}{d\omega d\Omega}\nonumber\\
&&\mspace{75mu}\times P^N_\textrm{part}(\ve p + \ve q,\omega-E-t_{A-1}),
\end{eqnarray}
where $E_{\ve p}=\sqrt{M^2+{\ve p}^2}$, $\sigma^\textrm{elem}_{\ell N}$ is the elementary cross section stripped off the energy-conserving $\delta$ function, and $t_{A-1}$ denotes the recoil energy of the residual nucleus, of mass $M_{A-1}=M_A-M+E$ and momentum $\ve p$.

The hole SF, $P^N_\textrm{hole}(\ve p,E)$, is the probability distribution of removing a nucleon $N$ with momentum $\ve p$ from the nuclear ground state, leaving the $(A-1)$-nucleon residual
system with excitation energy $E$, whereas the particle SF, $P^N_\textrm{part}(\ve{p'},\mathcal{T'})$, describes the propagation of the struck nucleon, carrying momentum $\ve{p'}$ and kinetic energy $\mathcal{T'}$.

The hole SF of carbon~\cite{ref:Omar_LDA}, used in this paper, has been obtained within the
local-density approximation (LDA), combining the information on the shell-model structure extracted from experimental data~\cite{ref:Saclay_C,ref:Dutta} with the correlation contribution calculated in uniform nuclear matter at different densities~\cite{ref:Omar_NM}. In addition to being extensively used in the analysis of electron-scattering data in a variety of kinematical regimes, the carbon SF of Ref.~\cite{ref:Omar_LDA} yields a momentum distribution consistent with the one  extracted from $(e,e'p)$ data at large missing energy and momentum~\cite{ref:Rohe}.

For the particle SF, we test two different approximations. In the crudest one~\cite{ref:Omar_oxygen}, Pauli blocking is accounted for through the action of the Heaviside
step function,  as in the Fermi gas model. The resulting expression is
\begin{equation}\label{eq:oldPB}
P^\theta_\textrm{part}(\ve{p'},\mathcal{T'})=\delta(\Epp-M-\mathcal{T'})\big[1-\theta(\overline p_F-\n{p'})\big],
\end{equation}
where $\overline p_F=211$ MeV is determined from the LDA average,
\[
\overline p_F=\int d^3r \rho(r) p_F(r),
\]
with $p_F(r)=\sqrt[3]{3\pi^2A\rho(r)/2}$. In the LDA treatment of Ref.~\cite{ref:shape}, the particle SF is calculated from the momentum distribution of isospin-symmetric nuclear matter
at uniform density $\rho$, $n^\textrm{NM}_\rho(\ve{p'})$, using
\begin{eqnarray}\label{eq:LDAPB}
P^\textrm{LDA}_\textrm{part}(\ve{p'},\mathcal{T'})&=&\delta(\Epp-M-\mathcal{T'})\nonumber\\
&&\times\left[1-\int d^3r\rho(r)C_\rho n^\textrm{NM}_\rho(\ve{p'})\right],
\end{eqnarray}
where $C_\rho=4\pi p^3_F(r)/3$. Note that $C_\rho n^\textrm{NM}_\rho(\ve{p'})$ corresponds to $\theta\big(p_F(r)-\n{p'}\big)$ in the local Fermi gas model.

To account for the distortion of the charged lepton kinematics arising from the interaction with the Coulomb field of the nucleus, we employ the effective momentum
approximation (EMA$'$), described in Ref.~\cite{ref:EMA'}.

\begin{figure*}
\centering
    \subfigure{\label{fig:electrons_a}}
    \subfigure{\label{fig:electrons_b}}
    \subfigure{\label{fig:electrons_c}}
    \subfigure{\label{fig:electrons_d}}
    \subfigure{\label{fig:electrons_e}}
    \subfigure{\label{fig:electrons_f}}
    \subfigure{\label{fig:electrons_g}}
    \subfigure{\label{fig:electrons_h}}
    \subfigure{\label{fig:electrons_i}}
    \includegraphics[width=0.85\textwidth]{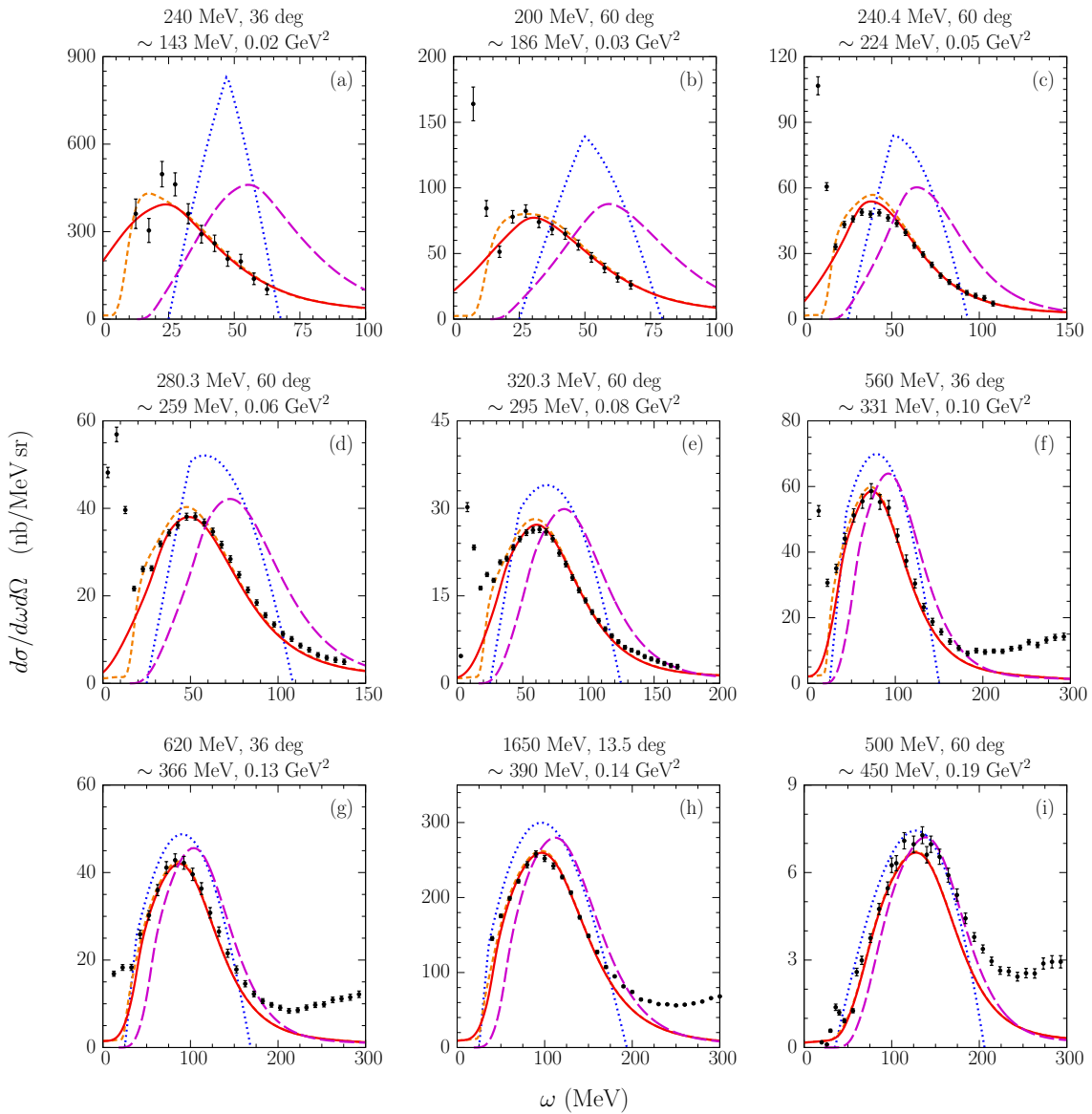}
\caption{\label{fig:electrons}(color online). Double differential electron-carbon cross sections, $\xsec$. The results obtained with Pauli blocking accounted for in the local-density (solid lines) and step-function (short-dashed lines) approximations are compared to the experimental data reported by (a)--(g) Barreau \etal{}~\cite{ref:Barreau}, (h) Baran \etal{}~\cite{ref:Baran}, and (i) Whitney \etal{}~\cite{ref:Whitney}. The IA (long-dashed lines) and
RFG calculations (dotted lines) are also shown, for reference. The panels are labeled according to beam energy, scattering angle, and values of $\n q$ and $Q^2$ at the quasielastic peak.
}
\end{figure*}

\begin{figure*}
\centering
    \includegraphics[width=0.80\textwidth]{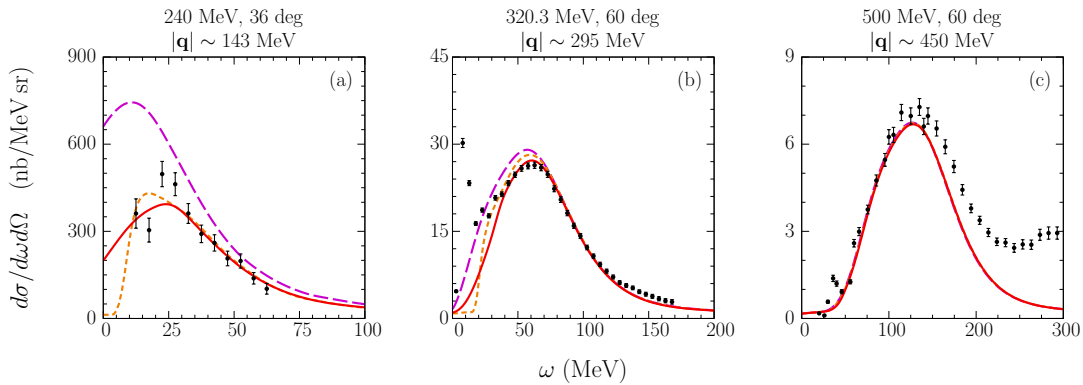}
\caption{\label{fig:PauliBlocking}(color online). Role of Pauli blocking at different kinematics, corresponding to the momentum transfers of (a) 143, (b) 295, and (c) 450 MeV at the quasielastic peak. The calculations neglecting this effect (long-dashed lines) are compared with those accounting for it in the local-density (solid lines) and step-function (short-dashed lines) approximations.
}
\end{figure*}

\begin{figure}
\centering
    \includegraphics[width=0.85\columnwidth]{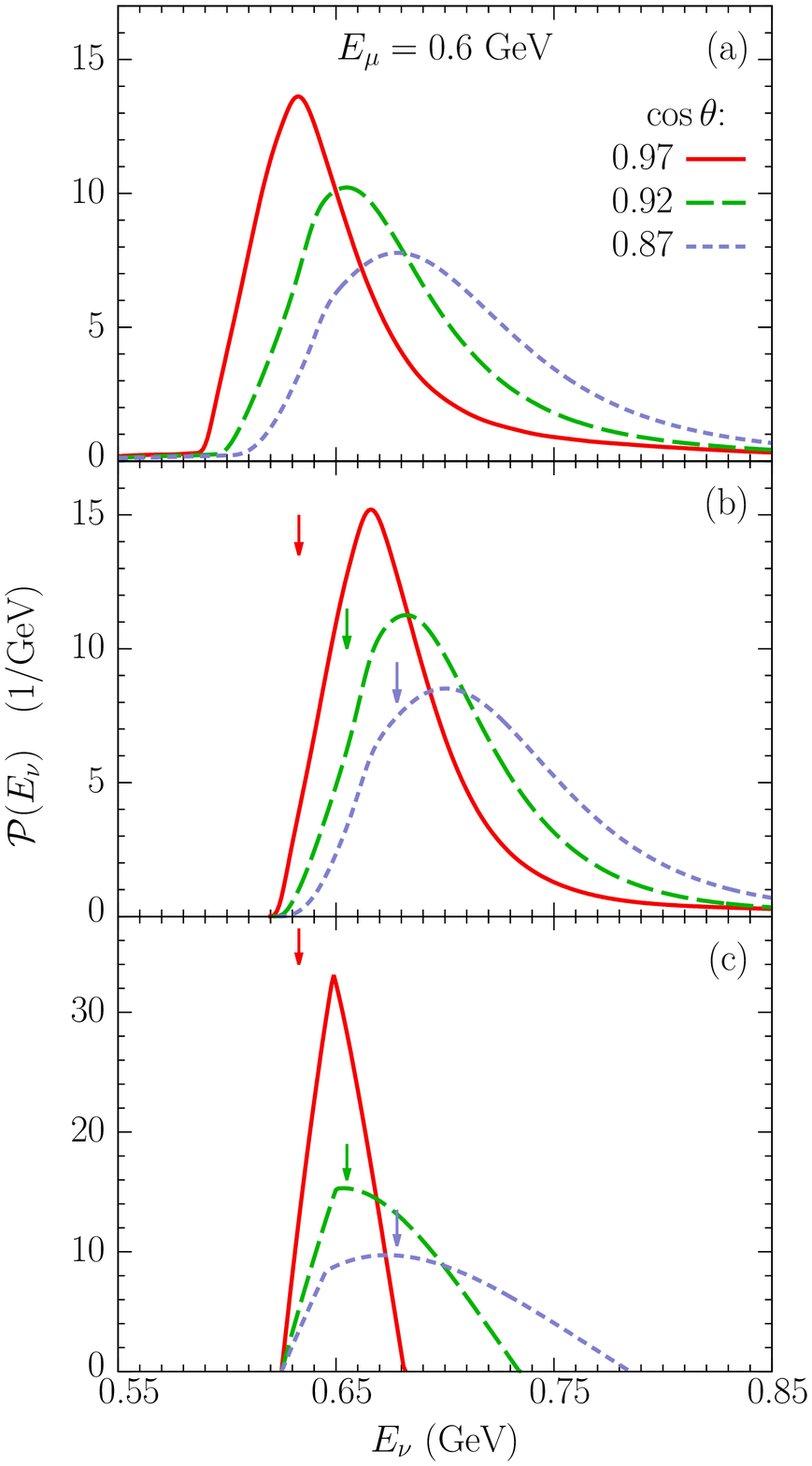}
\caption{\label{fig:enDistrib}(color online). Probability distributions for a muon of energy 0.6~GeV and cosine of production angle 0.97, 0.92, and 0.87 to originate from the interaction of a neutrino of energy $E_\nu$, calculated within (a) the approach discussed in this article, (b) the IA approach, and (c) the RFG model. The positions of maxima in (a), shown by arrows in (b) and (c), correspond to the $\n q$ ($Q^2$) values of approximately 156, 257, and 335 MeV (0.02, 0.06, and 0.11 GeV$^2$), respectively.
}
\end{figure}

\section{Numerical Results}\label{sec:NumericalResults}

In Fig.~\ref{fig:electrons},
the calculated $\isotope[12][6]{C}(e,e')$ cross sections are compared to the experimental data of Refs.~\cite{ref:Barreau,ref:Baran,ref:Whitney}.
Note that, in the carbon nucleus, the giant dipole resonance, of width $\Gamma_\textrm{c.m.} = 3.2$~MeV, can be excited providing the energy
$E_x=22.6$ MeV~\cite{ref:nuclearDataSheets_12,ref:tableOfIsotopes}. In addition, nuclear excitations belonging to the discrete spectrum are known to appear at
$E_x\lesssim10$ MeV~\cite{ref:Hofstadter,ref:Fregeau}. In the kinematics under consideration, the corresponding energy transfer is $\omega=E_x+t_A$, with the nuclear recoil energy $t_A$ increasing from 1.0~MeV in Fig.~\ref{fig:electrons_a}, to 4.5~MeV in Fig.~\ref{fig:electrons_e}, and to 10.9~MeV in Fig.~\ref{fig:electrons_i}.

The low-$\omega$ data also contain a contribution arising from elastic scattering on the nucleus, corresponding to $E_x=0$ MeV, well visible
in Figs.~\ref{fig:electrons_a}--\ref{fig:electrons_f}. This process is strongly quenched when the beam energy or the scattering angle increase, the cross section dropping down by 3 [2] orders of magnitude between the kinematics of Figs.~\ref{fig:electrons_b} and \ref{fig:electrons_i} [Figs.~\ref{fig:electrons_a} and \ref{fig:electrons_c}].

Having identified the contributions of the additional reaction mechanisms in the data, we can unambiguously analyze the QE peaks. The short-dashed and solid lines of
Fig.~\ref{fig:electrons},  representing the results obtained  using  the particle SF of Eqs. \eqref{eq:oldPB} and \eqref{eq:LDAPB}, respectively, clearly show that
our approach is capable to describe the measured cross sections with remarkable accuracy. Comparison to the long-dashed lines, corresponding to the IA results,
provides a measure of FSI effects, the inclusion of which turns out to be essential. In the kinematical conditions considered in our work, both FSI effects---the redistribution of the strength from the peak to the tails and the shift of the cross section toward lower energy transfers---play a significant role.  The real part of the proton optical potential,  driving the latter effect,
turns out to be $-33.3$, $-22.0$, and $-11.8$~MeV in the kinematical setups of Figs.~\ref{fig:electrons_a}, \ref{fig:electrons_e}, and \ref{fig:electrons_i}, respectively.

For comparison, in Fig.~\ref{fig:electrons}, we also show the results of RFG calculations performed with Fermi momentum 221~MeV and separation energy $\epsilon=25$ MeV. These parameters were determined in Ref.~\cite{ref:Whitney} from a fit to the data displayed in Fig.~\ref{fig:electrons_i}. It clearly appears that, while that data set is fairly well described by the RFG model, this is no longer
the case when the values of $\n q$ and $\omega$ (or $Q^2$) at the QE peak decrease. At the kinematics of Figs.~\ref{fig:electrons_h}--\ref{fig:electrons_a}, the calculated cross sections begin to
sizably disagree from the data and the inability of the RFG model to reproduce position, shape, and height of the QE peak becomes manifest. Even though for any~selected kinematical setup the observed discrepancies could be reduced adjusting the model parameters, such a procedure cannot bring the calculations to overall agreement with the $(e,e')$ data in the broad kinematical region of interest for neutrino oscillation studies.

Note that, thanks to the use of a realistic hole SF~\cite{ref:Omar_LDA}, our calculations account for some of the processes involving two-particle--two-hole ($2p2h$) final states, namely those triggered by
initial-state correlations between nucleons. This contribution produces the high-$\omega$ tail of the cross section, in excellent agreement with the data presented in Figs.~\ref{fig:electrons_a}--\ref{fig:electrons_c}. Nevertheless, at higher beam energy and scattering angle, the obtained cross sections tend to underestimate the experimental points above the QE peak, see Figs.~\ref{fig:electrons_d}--\ref{fig:electrons_i}.
As suggested in Ref.~\cite{ref:Donnelly_MEC}, this feature is likely to be related to the contribution of two-body reaction mechanisms~\cite{ref:Shimizu,ref:Alberico_MEC,ref:Dekker} not included in our approach, as well as to the occurrence of inelastic processes~\cite{ref:Nakamura}.

The authors of Ref.~\cite{ref:Alberico_MEC} correctly pointed out that $2p2h$ final states may have a threefold origin, resulting from one-body interactions
involving (i) initial- or (ii) final-state correlations, as well as from (iii) two-body  interactions, such as those involving meson-exchange currents.
However, their approach was based on a simplified description of nuclear dynamics in terms of perturbative pion exchange, in which correlation effects
were taken into account through an {\em ad hoc} modification of the $\pi NN$ vertex. The results of recent theoretical calculations carried out using state-of-the-art many-body wave functions
and current operators have confirmed the findings of the pioneering work of Ref.~\cite{ref:Alberico_MEC}, showing that the effect of  interference between the mechanisms leading to the
excitation of $2p2h$ final states are large, and must be taken into account in a consistent fashion~\cite{ref:Omar_2Body}. Therefore, the accurate description of the high-$\omega$ tail of the
QE peak is beyond the scope of the present work.

It is noteworthy that at low scattering angles, QE scattering may be the dominant reaction mechanism even for high beam energy. In such cases, our calculations are in good agreement with the data, as shown in Fig.~\ref{fig:electrons_h}.

To account for Pauli blocking, we apply two clearly different approximations, corresponding to Eqs. \eqref{eq:oldPB} and \eqref{eq:LDAPB}. It turns out, however, that in most cases discussed here, they yield rather similar results in the region of the QE peak. Where other reaction mechanisms do not contribute and a distinction can be made, the data show only a slight preference for the LDA prescription, see Figs.~\ref{fig:electrons_c}--\ref{fig:electrons_e}.

At low energy transfers, Pauli blocking plays an important role. Should it be neglected, the agreement with experimental cross sections would be spoiled, as shown in Fig.~\ref{fig:PauliBlocking}. The influence of Pauli blocking decreases when the momentum transfer becomes comparable with a typical nucleon momentum, of the order of $\overline p_F$, and vanishes at $\n q$ exceeding $2\overline p_F$. This behavior is well understood  within the RFG model.

Our approach can be readily applied to neutrino CC QE scattering.
To discuss how the procedure of energy unfolding is influenced by nuclear effects, we consider the probability distribution that a charged lepton of given energy $E_\ell$ and production angle $\theta$ originates from the interaction of a neutrino of energy $E_\nu$,
\begin{equation}\label{eq:P_of_E}
\mathcal P(E_\nu)\big|_{E_\ell,\:\cos\theta}= \frac{\frac{d\sigma(E_\nu)}{dE_\ell d\cos\theta}}{\int d E_\nu \frac{d\sigma(E_\nu)}{dE_\ell d\cos\theta}}.
\end{equation}

Figure~\ref{fig:enDistrib} shows a comparison between $\mathcal P(E_\nu)$ obtained from our calculations, the IA approach, and the RFG model, at the kinematics particularly relevant to $\nu_\mu$ CC inclusive interactions in the T2K experiment~\cite{ref:T2K_CC}. As this region corresponds to high values of the nuclear transparency, FSI do not cause a sizable broadening of $\mathcal P(E_\nu)$, leaving its shape and width largely unaffected. The main difference between the approach of this paper and the IA calculations is, therefore, related to the shift toward low $E_\nu$ produced by the real part of the optical potential.

On the other hand, the results of our approach and the RFG model turn out to clearly differ, in both shape and width. In the low-$E_\nu$ region, this behavior may be traced back to the different treatment of Pauli blocking. At high $E_\nu$, it is a consequence of nucleon-nucleon correlations, neglected in the RFG model, which lead to the appearance of the tail of the cross section in our calculations.

As for the neutrino energies corresponding to the maximum of $\mathcal P(E_\nu)$, our approach and the RFG model give rather consistent results for $\cos\theta$ of 0.92 (0.87), the corresponding values being  655 and 653 (678 and 673) MeV, respectively. However,  at $\cos\theta=0.97$, we observe a larger difference between the energies predicted by our approach, 633 MeV, and the RFG model, 649 MeV. Should the separation energy 34 MeV~\cite{ref:MiniB_kappa} be applied in the RFG calculations instead of 25 MeV, the maxima would shift by $\sim$9 MeV, further increasing the discrepancy.

The observed differences do not come as a surprise, in view of the $(e,e')$ cross sections discussed above. Note that in the kinematical conditions corresponding to the maxima in Fig.~\ref{fig:enDistrib} and to the QE peaks of Figs.~\ref{fig:electrons_a}, \ref{fig:electrons_d}, and \ref{fig:electrons_f} the nuclear response is probed in a similar region of the $\big(\n q, Q^2\big)$ plane.
The width and the position of the QE peak in electron scattering are determined by the same nuclear effects which shape the neutrino CC QE cross section in Eq.~\eqref{eq:P_of_E}. Therefore, the large body of available $\isotope[12][6]{C}(e,e')$ data clearly shows that accurate energy unfolding, required by precise oscillation measurements, cannot be performed using the RFG model. On the other hand, our approach reproduces those data well, which allows us to expect a comparable accuracy in neutrino interactions.

The double differential CC QE cross section entering the definition of $\mathcal P(E_\nu)$ is obviously affected by processes involving the excitation of $2p2h$ final states. As their contribution cannot be currently taken into account in an {\it ab initio} manner, one needs to resort to an effective approach.

A~consistent analysis of the NC and CC QE cross sections measured by the BNL E734~\cite{ref:BNL_E734_NC}, MiniBooNE~\cite{ref:MiniB_CC,ref:MiniB_NC}, and NOMAD~\cite{ref:NOMAD} experiments, performed in Ref.~\cite{ref:carbon}, leads to the observation that---barring the normalization of the MiniBooNE cross sections---the available data are in agreement with the results of SF calculations carried out setting the axial mass to $M_A=1.23$ GeV, a value larger than the
one extracted from deuteron measurements. Because it applies to both the NC QE cross sections as a function of knocked-out proton's energy and the CC QE cross sections
depending on charged lepton's kinematics, this finding seems to point to a~general property of the cross section in the SF approach.

Therefore, the calculations presented in Fig.~\ref{fig:enDistrib} are performed using the axial mass $M_A=1.23$ GeV~\cite{ref:MiniB_kappa}. However, setting $M_A=1.03$ GeV~\cite{ref:Meissner} would change $\mathcal P(E_\nu)$ at the peak by less than 0.50 (0.85)\% for $\cos\theta$ of 0.97 (0.87). Such a weak dependence on the axial mass follows from the definition of $\mathcal P(E_\nu)$, independent of the cross section's absolute normalization, and from the rather narrow range of $Q^2$ yielding a sizable contribution to the discussed results.

We observe that for $E_\nu\sim600$ MeV, the Coulomb energies of the struck nucleon and the final lepton, together with a~somewhat deeper binding of neutrons than protons, introduce a sizable difference between neutrino and antineutrino CC QE scattering. This effect is of utmost importance for the measurements of CP violation. For a 600-MeV antimuon observed at $\cos\theta$ of 0.97 (0.87), we find the most probable $\bar\nu_\mu$ energy lower by 14 (17) MeV than that of $\nu_\mu$ for muon at the same kinematics.

The relevance of the above observation for oscillation experiments can be illustrated considering reconstructed energy distributions.
Recall that the standard reconstruction method is based on the observation that in CC QE  scattering off a~bound nucleon at rest, the energy can be exactly determined from the measured kinematics of the charged lepton only. The energy and momentum conservation than give
\begin{equation}\label{eq:recE}
E_\nu^\textrm{($N$ at rest)}= \frac{2E_\ell\tilde M-(m^2+\tilde M^2 - M^2)}{2 ( \tilde M - E_\ell + \n{k_\ell}\cos\theta )},
\end{equation}
where $\tilde M=M -\epsilon$ and $\ve{k}_\ell^2=E_\ell^2-m^2$, with $\epsilon$ and $m$ being the nucleon separation energy and the mass of the charged lepton, respectively. The same expression applied to CC QE scattering off a~nuclear target,
\begin{equation}\label{eq:standardReconstruction}
\Erec= E_\nu^\textrm{($N$ at rest)},
\end{equation}
defines the reconstructed energy~\cite{ref:K2K_PRL,ref:shape,ref:Martini_Erec,ref:Nieves_Erec,ref:Lalakulich_Erec}.

\begin{figure}
\centering
    \includegraphics[width=0.85\columnwidth]{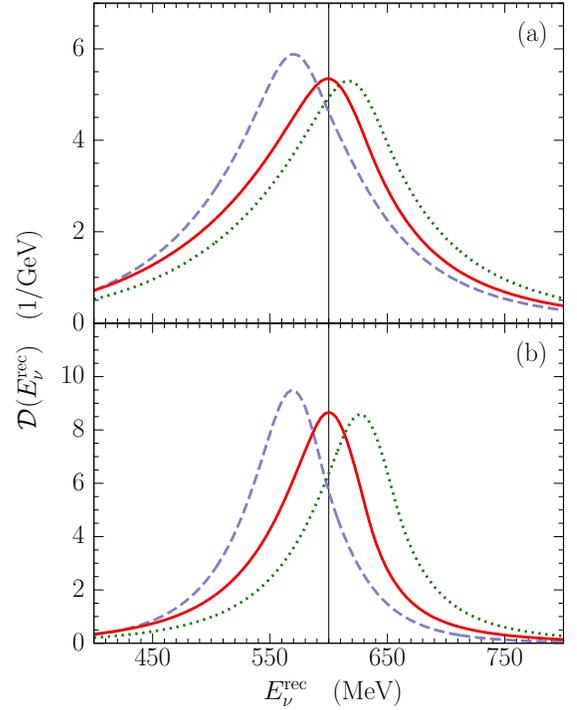}
\caption{\label{fig:stdRec}(color online). Relevance of FSI effects on the reconstructed energy distributions in (a) $\nu_\mu$ and (b) $\bar\nu_\mu$ interactions at the true energy 600 MeV. Our calculations (solid lines) are compared to the IA results (dashed lines), setting the separation energy to 19 MeV [6 MeV] in (a) [(b)]. For reference, our calculation with $\epsilon$ of 34  MeV [30 MeV] (dotted lines) is also shown in (a) [(b)].
}
\end{figure}

To determine the parameter $\epsilon$, we require the $\Erec$ distribution to be peaked at the true energy $E_\nu=600$ MeV. The corresponding value of the separation energy is 19 (6) MeV in the case of $\nu_\mu$ ($\bar\nu_\mu$) scattering described by the approach of this article.

As shown in Fig.~\ref{fig:stdRec}, the effects of FSI on the $\Erec$ distributions calculated at $E_\nu=600$ MeV are a $\sim$30-MeV shift of the maximum, produced by the real part of the optical potential, and a broadening of the distribution, resulting from the folding function.
We also observe that for $\epsilon$ fixed to 34 (30) MeV in $\nu_\mu$ ($\bar\nu_\mu$) energy reconstruction as in Refs.~\cite{ref:MiniB_kappa,ref:MiniB_anu}, the maximum of the $\Erec$ distribution in our calculations is off by 17 (27) MeV.

Note that the shift of the $\Erec$ distribution is larger than the average value of $U_V$ which drives this effect. The reason is twofold: (i) the dependence of $\Erec$ on the energy
of the charged lepton is nonlinear, see Eq.~\eqref{eq:recE}, and (ii) it is minimal at $\cos\theta=1$, where $E_\ell$ is subject to the largest change by the real part of the optical potential.

%%%%%%%%%%%%%%%%
\begin{table}
\caption{\label{tab:maxima} Positions of the maximum of the $\Erec$ distribution as a function of the true neutrino energy, in units of MeV.
The results of the approach proposed in this article (FSI) are compared to those obtained within the IA, for different values of the separation energy entering the  definition
of $\Erec$ [Eq.~\eqref{eq:recE}]. The numerical uncertainty is estimated to be 1~MeV.
}
\begin{ruledtabular}
    \begin{tabular}{@{}lrrrrr@{}}
    $E_\nu$ & {200} & {400} & {600} & {800} & {1000}\\[3pt]
    \hline
    FSI, $\nu_\mu$, $\epsilon = 19$ MeV& {211} & {401} & {600} & {799} & {998}\\
    IA, $\nu_\mu$, $\epsilon = 19$ MeV& {173} & {370} & {570} & {770} & {970}\\
    FSI, $\nu_\mu$, $\epsilon = 34$ MeV & {229} & {419} & {617} & {816} & {1015}\\
    \hline
    FSI, $\bar\nu_\mu$, $\epsilon = 6$ MeV& {210} & {402} & {600} & {799} & {999}\\
    IA, $\bar\nu_\mu$, $\epsilon = 6$ MeV& {172} & {369} & {569} & {769} & {969}\\
    FSI, $\bar\nu_\mu$, $\epsilon = 30$ MeV& {239} & {429} & {627} & {826} & {1025}\\
    \hline
    FSI, $\nu_e$, $\epsilon = 19$ MeV& {206} & {401} & {599} & {799} & {998}\\
    FSI, $\bar\nu_e$, $\epsilon = 6$ MeV& {206} & {402} & {600} & {799} & {999}\\
    \end{tabular}
\end{ruledtabular}
\end{table}
%%%%%%%%%%%%%%%%

At the kinematics most relevant to the T2K experiment, the observed shift of the maximum does not show a sizable $E_\nu$ dependence. Table~\ref{tab:maxima} illustrates that the separation energies determined at $E_\nu=600$ MeV are able to bring the peaks of the $\Erec$ distributions into a good agreement with the true values of (anti)neutrino energy, and the differences do not exceed 2 MeV for $E_\nu$ between 400 and 1000 MeV. This is also the case for interactions of electron neutrino and antineutrino.

According to our findings, the separation energies in neutrino and antineutrino scattering differ by 13 MeV, compared to 4 MeV assumed in the analysis of the MiniBooNE~\cite{ref:MiniB_kappa,ref:MiniB_anu} and MINER$\nu$A~\cite{ref:MINERvA_anu,ref:MINERvA_nu} experiments. While we estimate that the Coulomb field of the carbon nucleus changes the energy of a~charged particle by 3.5 MeV, this modification affects the $\ell^-$ and $\ell^+$ energies with opposite sign, and adds to the difference between the real part of the optical potential for proton and neutron. Moreover, Coulomb effects shift the neutron and proton energy levels in carbon by 2.8 MeV~\cite{ref:carbon}.
Although each of those modifications is small, altogether they introduce a~sizable difference between neutrino and antineutrino interactions, $3\times3.5+2.8\approx13$ MeV, that will be crucial in the context of measurements of CP violation.

\begin{figure}
\centering
    \subfigure{\label{fig:newRec_a}}
    \subfigure{\label{fig:newRec_b}}
    \subfigure{\label{fig:newRec_c}}
    \includegraphics[width=0.85\columnwidth]{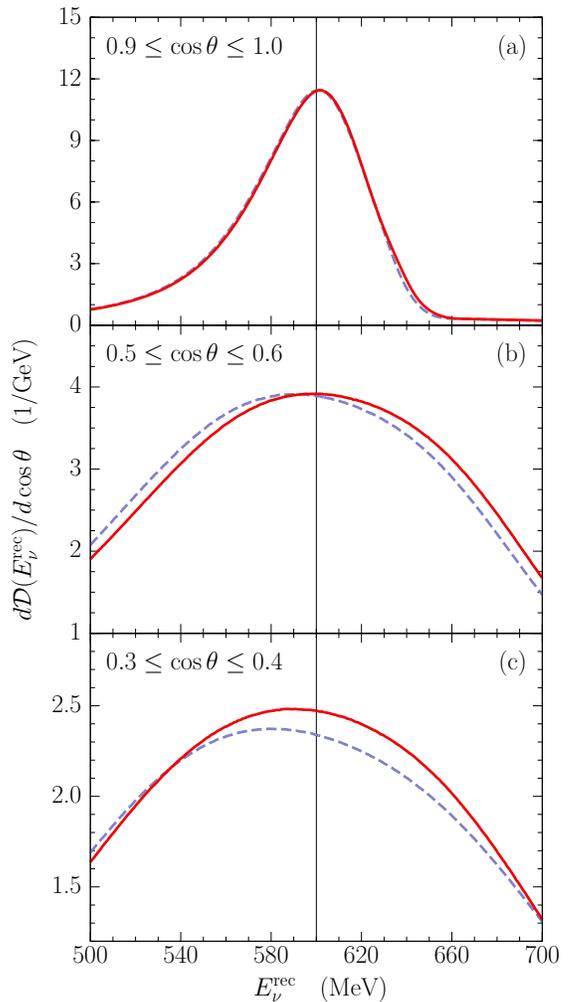}
\caption{\label{fig:newRec}(color online). Contributions to the $\Erec$ distribution coming from (a) $0.9\leq\cos\theta\leq1.0$, (b) $0.5\leq\cos\theta\leq0.6$, and (c) $0.3\leq\cos\theta\leq0.4$ bins, calculated at the true energy of 600 MeV. The new method of neutrino energy reconstruction [Eq.~\eqref{eq:newReconstruction}, solid lines]
is compared with the standard one [Eq.~\eqref{eq:standardReconstruction}, dashed lines].
}
\end{figure}

Analyzing electron-scattering data in Fig.~\ref{fig:electrons}, we have observed that a~constant value of the separation energy is not able to reproduce the position of the QE peak over the broad kinematic region relevant to neutrino oscillation studies.
Because of the universality of nuclear effects affecting electron and neutrino interactions, the same issue has reappeared in the context of the probability distributions $\mathcal P(E_\nu)$, shown in Fig.~\ref{fig:enDistrib}. However, the energy reconstruction by means of Eq.~\eqref{eq:standardReconstruction} treats the separation energy as a~single parameter.

To improve the accuracy of energy reconstruction, one should also take into account, in Eq.~\eqref{eq:recE}, the dependence of separation energy on
the charged lepton's kinematics, described by the function \[\epsilon = \epsilon(E_\ell,\:\cos\theta).\]

The above function can be determined, for example, requiring the positions of the maximum of $\mathcal P(E_\nu)$ to be reproduced over the kinematics of interest.

Equivalently, energy reconstruction can be performed using the information on the maxima of $\mathcal P(E_\nu)$ in a direct way, identifying the most probable
neutrino energy at $E_\ell$ and $\cos\theta$ with the reconstructed energy,
\begin{equation}\label{eq:newReconstruction}
\Erec= \Emax,
\end{equation}
where $\Emax = \Emax(E_\ell,\:\cos\theta)$ is the energy corresponding to the maximum of $\mathcal P(E_\nu)$ at $E_\ell$ and $\cos\theta$.

To illustrate the difference between this novel method of energy reconstruction and the standard one, based on Eq.~\eqref{eq:standardReconstruction}, in Fig.~\ref{fig:newRec} we compare the corresponding contributions to the $\Erec$ distribution coming from different $\cos\theta$ bins, calculated for the true energy $E_\nu=600$ MeV. At $0.9\leq\cos\theta\leq1.0$, the standard and new methods are in good agreement, see Fig.~\ref{fig:newRec_a}, yielding distributions peaked at 601 and 602 MeV, respectively. When the muon production angle increases, we observe an increasing difference between the true value of energy and the peak of the standard $\Erec$ distribution, located at 591 and 581 MeV for $0.5\leq\cos\theta\leq0.6$ and $0.3\leq\cos\theta\leq0.4$, respectively. On the other hand, the maxima predicted by the new reconstruction method, 599 and 591 MeV
in Figs.~\ref{fig:newRec_b} and \ref{fig:newRec_c}, respectively, turn out to be in better agreement with $E_\nu$.

The decreasing accuracy of the the standard reconstruction method at higher production angles can be traced back to the real part of the optical potential, which in our approach is directly related to $\cos\theta$, see Eq.~\eqref{eq:t}. For example, over the interval $0.9\leq\cos\theta\leq1.0$ $(0.3\leq\cos\theta\leq0.4)$ the $\Erec$ distribution for $E_\nu=600$ MeV picks up contributions corresponding to the real part of the optical potential $-38\leq U_V \leq-25$ MeV ($-6\leq U_V \leq-5$ MeV). The separation energy fixed to reproduce the true energy at the dominant kinematics ($\cos\theta\sim0.83$) turns out not to be appropriate at the subdominant ones. Note, however, that the kinematics of Figs.~\ref{fig:newRec_a}, \ref{fig:newRec_b}, and \ref{fig:newRec_c} play non-negligible role in CC QE scattering, contributing 8.2, 8.2, and 6.2\% of the cross section, respectively, compared to the dominant 10.9\% contribution from $0.8\leq\cos\theta\leq0.9$.

The new method of energy reconstruction, which, by construction, accounts for the shift produced by the real part of the optical potential, appears to be able to bring the reconstructed energy into better agreement with its true value.

\section{Conclusions}\label{sec:Conclusions}

We have carried out a systematic study of the electroweak response of carbon in the QE sector. Our approach, based on a generalization of the SF formalism, allows for a consistent
inclusion of FSI effects and does not involve any adjustable parameters.

To validate our computational scheme and quantitatively assess its accuracy, in view of applications to the study of neutrino interactions, we have performed an extensive comparison to the available electron-scattering data \cite{ref:Barreau,ref:Whitney,ref:QES_archive,ref:OConnell,ref:Baran,ref:Baghdasarian,ref:Sealock,ref:Day}.
%to draw reliable conclusions on uncertainties of the present calculations.
These results, presented in this paper and provided as Supplemental Material~\cite{ref:SupplMat}, allow us to assign a 5\% uncertainty to the absolute values of the $(e,e')$ cross sections, in the absence of interaction mechanisms other than QE scattering induced by one-nucleon currents. In the region of $\n q$ and $\omega$ (or, equivalently, $Q^2$) under consideration, we estimate at 5 MeV the uncertainty of the QE peak position, consistently with the value deduced from comparisons of different parametrizations of the carbon optical potential~\cite{ref:Cooper_EDAI,ref:Cooper_dem}.  In the kinematical setups corresponding to the excitation energy $E_x\lesssim26$ MeV, where the basic assumptions of our model no longer apply, we assign 100\% uncertainty to the obtained results. We emphasize that this is a general limitation of the IA framework, and not a~feature specific to the model of this article.

Our approach allows a consistent and accurate determination of the neutrino and antineutrino cross sections, which will be required for the measurement of the CP violating phase.

In the considered kinematical conditions, we assign to $\mathcal P(E_\nu)$ a conservative 2\% uncertainty related to the axial mass value. Because the assumptions underlying our approach do not apply to scattering at excitation energy below $\sim$26 MeV, and in this region we are not able to validate our calculations against data, the obtained results for $\cos\theta=0.97$ (0.87) should be considered 100\% uncertain for $E_\nu\lesssim631$ (634) MeV. We estimate that the most probable values of $E_\nu$ resulting from our calculations are determined with an uncertainty of 5 MeV.

Analyzing the standard method of neutrino energy reconstruction, we have found that the values of the separation energy suitable at the kinematics of the T2K experiment are 19 MeV for neutrinos and 6 MeV for antineutrinos. We estimate their (correlated) uncertainties, coming predominantly from the uncertainty of the carbon optical potential, at 6 MeV. The 13-MeV difference between the separation energies in the neutrino and antineutrino case, that will play a critical role in the searches for CP violation, is subject to an uncertainty of 2 MeV.

Based on the comparison between different production angles at fixed neutrino energy, we have argued that the standard reconstruction method cannot be accurate over a broad kinematical region. As a remedy, we have proposed a new procedure, exploiting the precise determination of the QE peak position provided by our approach. It may be used to increase the accuracy of energy reconstruction in experiments collecting events in a broad range of production angles.

The availability of accurate and reliable calculations of the cross section over the broad kinematical region relevant to the flux-integrated neutrino cross sections allowed for a
trustable analysis of the reconstruction of neutrino energy, the results of which clearly show that the unfolding procedure is significantly affected by the description of nuclear structure and dynamics.

The approach based on nuclear many-body theory and the spectral function formalism has reached a remarkable degree of accuracy in describing processes induced by the one-nucleon current. We hope that, thanks to the provided uncertainty estimates, it will turn out to be useful for analysis of experimental data, and that uncertainty estimates will become a new standard in theoretical modeling of neutrino cross sections. The extension of our approach outlined in Ref.~\cite{ref:Omar_2Body}, allowing for a consistent treatment of processes involving the excitation of $2p2h$ final states, will be an important step towards the development of an accurate, self-consistent, and complete description of neutrino cross sections. As advocated by the authors of Ref.~\cite{ref:Coletti}, such an approach is necessary for a reliable analysis of precise oscillation experiments performed over a broad kinematical region, where different reaction mechanisms contribute.

As a final remark, we note that in this analysis, we have not considered QE-like neutrino interaction arising from resonance excitation followed by meson absorption by the nucleus. As this mechanism constitutes an important irreducible background to QE scattering~\cite{ref:Leitner}, we leave its consistent inclusion for future studies.

\begin{acknowledgments}
We would like to thank Shinichi Hama for providing us with the parametrization of carbon optical potential in the \textsc{global} code. A.M. A. was supported by JSPS under Grant No. PE 13056. The work of O. B. was supported by INFN through grant MB31. The work of A.M. A. and M. S. is partly supported by JSPS under the Grant-in-Aid for Scientific Research (B) No. 23340073.
\end{acknowledgments}

\end{document}